\documentclass{PoS}

\PoS{PoS(LAT2005)333}
\graphicspath{{./Figures/}}                           

\newcommand{\vs}{\textit{\mbox{vs.\ }}}               
\renewcommand{\Re}{\mathfrak{Re}\,}                   
\newcommand{\Tr}{\mbox{Tr}}                           
\newcommand{\bc}{\texttt{bc}}                         
\newcommand{\fc}{\texttt{fc}}                         
\newcommand{\Ncp}{N_{\rm cp}{}}                       
\newcommand{\Fig}[1]{Fig.~\ref{#1}}

\newcommand{\Eq}[1]{Eq.~(\ref{#1})}

\newcommand{\preprint}{\newline%
  \begin{picture}(0,0)
  \put(293,100){\rm\small HU--EP--05/52, LU-ITP 2005/021}
  \end{picture}}

\title{Studying the infrared region in Landau gauge QCD\thanks{This work 
has been supported by the DFG under contract FOR~465.}\preprint} 

\ShortTitle{Studying the infrared region in Landau gauge QCD}

\author{A.~Sternbeck\thanks{Speaker. Supported by the DFG-funded
    graduate school GK~271.}\\ 
  Humboldt-Universit\"at zu Berlin, Institut f\"ur Physik,
  D-12489 Berlin, Germany\\
  E-mail: \email{andre.sternbeck@physik.hu-berlin.de}
}

\author{E.--M.~Ilgenfritz\\
  Humboldt-Universit\"at zu Berlin, Institut f\"ur Physik,
  D-12489 Berlin, Germany\\
  E-mail: \email{ilgenfri@physik.hu-berlin.de}
}

\author{M.~M\"uller-Preussker\\
  Humboldt-Universit\"at zu Berlin, Institut f\"ur Physik,
  D-12489 Berlin, Germany\\
  E-mail: \email{mmp@physik.hu-berlin.de}
}

\author{A.~Schiller\\
  Universit\"at Leipzig, Institut f\"ur Theoretische 
  Physik, D-04109 Leipzig, Germany\\
  E-mail: \email{Arwed.Schiller@itp.uni-leipzig.de}
}

\abstract{We report on the progress we made in studying the infrared behavior
of the ghost and gluon dressing functions in Landau gauge. Related to
this we also investigate a running coupling given in terms of
those functions and compare our results to those coming from the
Dyson-Schwinger approach. We present first numerical results for the $SU(3)$
ghost-ghost-gluon vertex renormalization constant. In addition the spectrum of
low-lying eigenvalues and eigenfunctions of the Faddeev-Popov operator
is determined. The saturation of the ghost propagator in terms of
those eigenvalues and eigenmodes is discussed at lower momenta.}

\FullConference{XXIIIrd International Symposium on Lattice Field Theory\\
         25-30 July 2005\\
         Trinity College, Dublin, Ireland}

\begin{document}

\section{Introduction}

The predicted infrared behavior of both the gluon and ghost propagator in QCD
was the starting point of our combined investigation of those Green's functions.
In fact, solutions of a truncated set of Dyson-Schwinger equations
(DSE) for both propagators in Landau gauge predict 
their dressing functions $Z_{D}$ and $Z_{G}$ to follow the power laws 
\cite{vonSmekal:1997isvonSmekal:1998Alkofer:2000wg,Lerche:2002ep,
Zwanziger:2001kw} 
\begin{equation}
   Z_{D}(q^2) \propto (q^2)^{2\kappa} \quad\textrm{and} 
          \quad Z_{G}(q^2) \propto (q^2)^{-\kappa}\quad
          \textrm{with} \quad \kappa\approx 0.595 
  \label{eq:infrared-behavior}
\end{equation}
at (extremely) low momentum. Thus, an infrared vanishing
gluon propagator occurs in intimate connection with a diverging
ghost propagator. The latter would be in agreement with the
Zwanziger-Gribov horizon condition as well as with 
the Kugo-Ojima confinement criterion
\cite{Zwanziger:2003cf,Zwanziger:1993dhGribov:1977wmKugo:1979gm}. 

Furthermore, in terms of the gluon and ghost dressing functions
a running coupling $\alpha_s(q^2)$ with a momentum dependence 
\begin{equation}
  \label{eq:runcoupling}
  \alpha_s(q^2) = \frac{g^2_0}{4\pi}\, Z^2_{G}(q^2)\,Z_{D}(q^2)
  \qquad\textrm{with}\ g_0^2 = 6/\beta\ \textrm{for}\ SU(3)
\end{equation}
can be defined. This construction is based on the consideration of the
ghost-ghost-gluon vertex where the renormalization function $Z_1$ of
this vertex is assumed to be constant at all momenta. Given the
asymptotic behavior in \Eq{eq:infrared-behavior} the running coupling
would have a finite limit at zero momentum \cite{Lerche:2002ep}.

Those infrared properties are claimed to be independent of how the
DSEs are truncated. However, it is worthwile to cross-check them using
lattice simulations \cite{Bloch:2002we}. We have been pursuing this
for the $SU(3)$ case during the last two
years \cite{Sternbeck:2005tk}. In addition, we have
begun to study the ghost-gluon vertex and the spectral properties of
the Faddeev-Popov operator. The low-lying eigenmodes of the latter are
believed to be immediately reflected by a diverging ghost propagator.

\section{The dressing functions and the running coupling} 

We have simultaneously studied the gluon and ghost propagators in the
quenched approximation. Thermalized $SU(3)$ gauge field configurations
$U=\{U_{x,\mu}\}$ have been put into the Landau gauge.
It is well-known that the Landau gauge functional 
\begin{equation}
  F_{U}[g] = \frac{1}{4V}\sum_{x}\sum_{\mu=1}^{4}\Re\Tr
  \;{}^{g}U_{x,\mu}
  \qquad\textrm{with}\quad g_x\in SU(3)
  \label{eq:functional}
\end{equation}
has several local maxima (Gribov copies), each satisfying the lattice 
Landau gauge condition $\partial_{\mu} A_{\mu}=0$. To explore to what extent 
this ambiguity has a significant influence on gauge
dependent observables, we have gauge-fixed each thermalized configuration
$\Ncp$ times using the \emph{over-relaxation} algorithm, starting from
a random gauge copy. For each configuration $U$, we have selected the
first (\fc) and the best (\bc) gauge copy (that with the largest
functional value) for subsequent measurements. For details we refer to
our recent publication \cite{Sternbeck:2005tk}. 

It turns out that the Gribov ambiguity has no systematic influence on
the infrared behavior of the gluon propagator. In marked contrast to
this, the ghost propagator at lower momenta depends on the selection of
gauge copies. This can be seen in \Fig{fig:ghost} where in the upper 
panel the ghost dressing function $Z_{G}$ is presented 
as a function of momentum $q$. The lower panel shows the ratio of the
corresponding functions measured on \fc{} and \bc{} gauge copies. 
We observe that the influence of the Gribov ambiguity becomes 
important at momenta $q<1$~GeV. However, the ratio for $\beta=6.0$ is
larger than the corresponding ratio for $\beta=5.8$ in the sensitive
momentum region at the same lattice size $24^4$.
Therefore, we conclude that the systematic Gribov copy effect
at the same momentum is reduced if the physical volume is increased.

\begin{figure}[b]
  \centering
  \includegraphics[height=5cm]{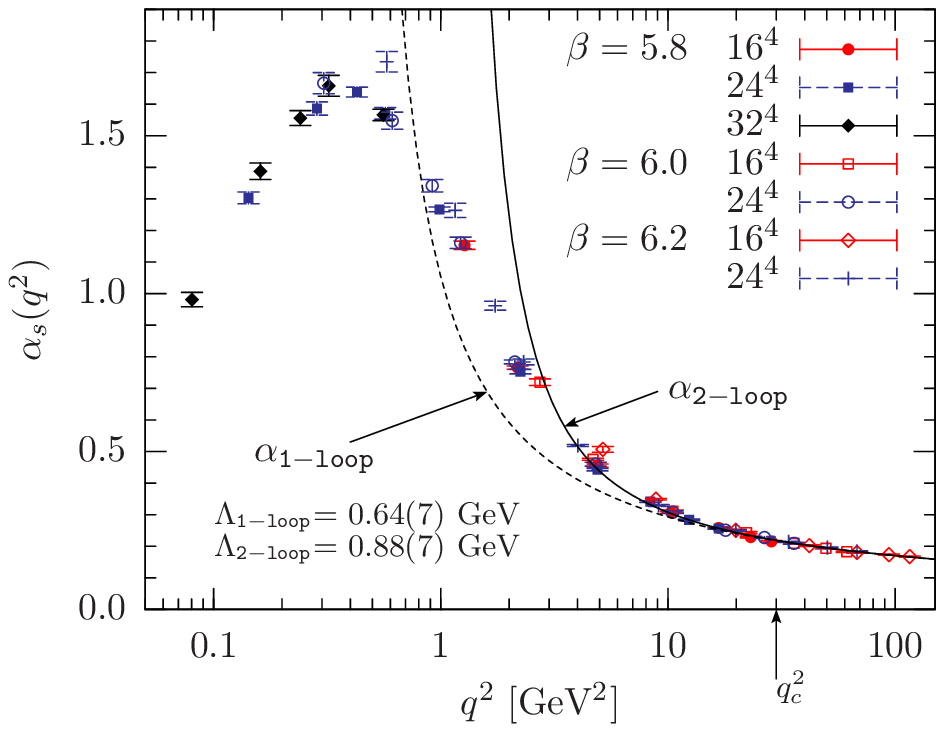}\quad 
  \includegraphics[height=5cm]{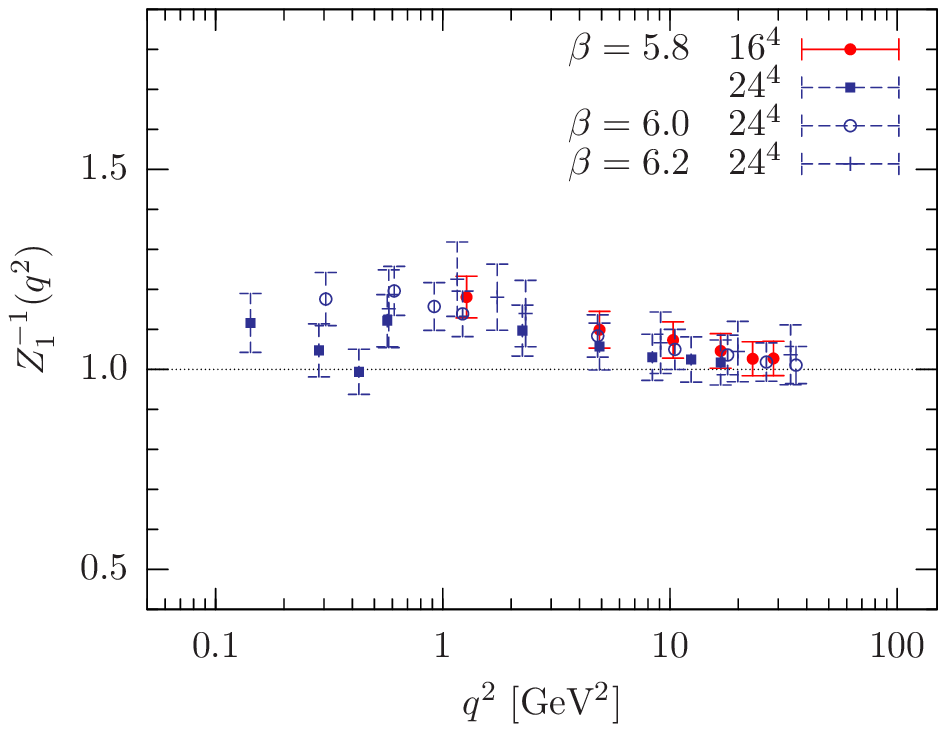}
  \caption{The momentum dependence of the running coupling
    $\alpha_{s}(q^2)$ (left) and of the ghost-ghost-gluon vertex
    renormalization function $Z_1(q^2)$ is shown. 
    Both have been measured on \bc{} gauge copies.} 
\label{fig:alpha_vertex}
\end{figure}

\begin{floatingfigure}[r]
  \includegraphics[width=7cm]{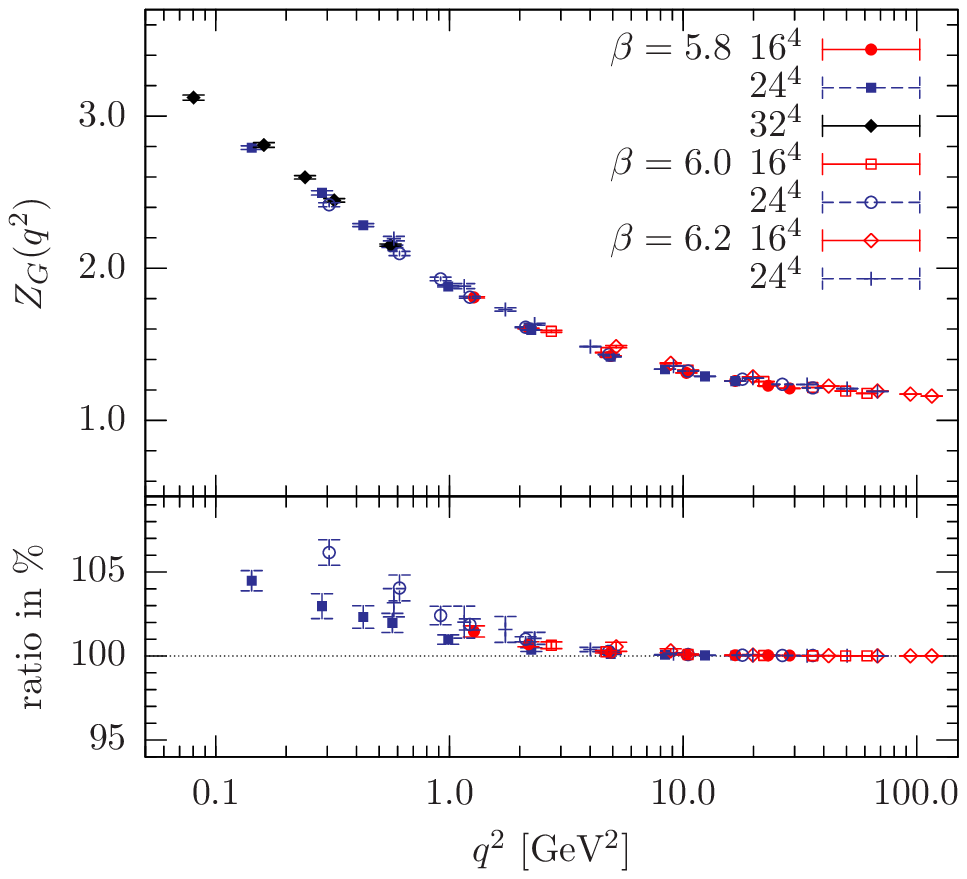}
\vspace{-0.3cm} 
  \caption{The ghost dressing function $Z_{G}$ measured on
    \bc{} gauge copies is shown \vs the (physical)
    momentum $q^2$. The lower panel shows the ratio
    $\langle Z^{\rm fc}\rangle / \langle Z^{\rm bc}\rangle$
    of dressing functions determined on first (\fc{}) and best (\bc{})
    gauge copies.}
\label{fig:ghost}
\end{floatingfigure}
We have first studied the infrared behavior of the gluon and ghost
dressing functions. In particular, we have tried to confirm the proposed
power laws \Eq{eq:infrared-behavior}. However, fitting both functions
for $q<0.5$~GeV we were unable to extract a common exponent 
\mbox{$\kappa>0.5$}.\footnote{We should mention that the number of data points
in this momentum region is certainly too small to fit the asymptotic
behavior for $q^2 \to 0$} This situation is also reflected in the
infrared behavior of the running coupling $\alpha_s(q^2)$ which is
proportional to the product $Z^2_{G}(q^2)Z_{D}(q^2)$ of 
the two dressing functions (cf.\ the left hand side of
\Fig{fig:alpha_vertex}). There the fits to the 1-loop and 2-loop 
running coupling are also shown.

One clearly sees that the running coupling monotonously increases with
decreasing momentum as long as \mbox{$q^2>0.4$~GeV$^2$}. 
After passing a maximum $\alpha_s(q^2)$ decreases. 
One could blame finite volume effects for this disagreement with
the continuum predictions of the DSE approach. Indeed, recent DSE investigations
\cite{Fischer:2005Fischer:2002eqFischer:2002hnFischer:2005bc} on a
torus also show a modified infrared behavior of the gluon and ghost dressing
functions.
One could also question whether the ghost-ghost-gluon
vertex renormalization function $Z_1(q^2)$ is really constant at lower
momenta. A recent investigation dedicated to the
ghost-ghost-gluon-vertex renormalization function $Z_1(q^2)$  
for the case of $SU(2)$ gluodynamics \cite{Cucchieri:2004sq} supports 
that \mbox{$Z_1(q^2)\approx 1$} at least for momenta larger than 1~GeV. 
We have performed a similar study of $Z_1(q^2)$ for the case of
$SU(3)$. Our (preliminary) results are presented on the right hand side of
\Fig{fig:alpha_vertex}. So far the data do not allow 
to draw a final conclusion whether $Z_1(q^2)$ stays constant at lower
momenta.

\section{Infrared properties of the Faddeev-Popov operator}

\begin{floatingfigure}[l]
  \includegraphics[width=7cm]{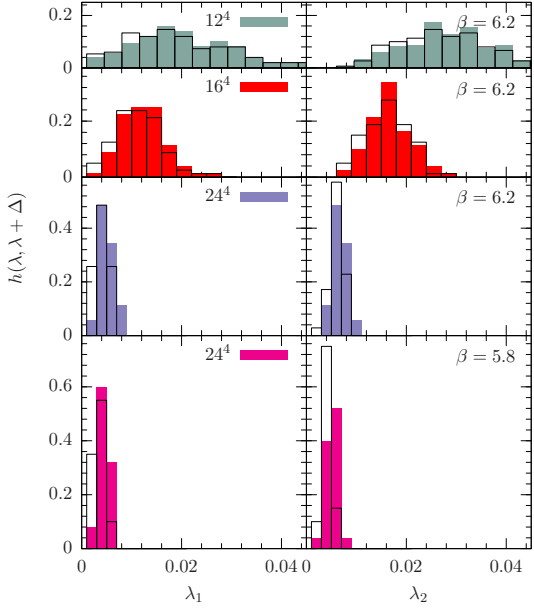}
  \caption{The distribution $h(\lambda)$ of the lowest
    (left panels) and second lowest (right panels) Faddeev-Popov 
    eigenvalues $\lambda_1$ and $\lambda_2$. Filled columns refer to 
    \bc{} gauge copies, while open columns refer to \fc{} copies.}
  \label{fig:fps_lowest}
\end{floatingfigure}
Next we discuss the spectrum of the low-lying eigenvalues
$\lambda_i$ of the Faddeev-Popov \mbox{(F-P)} operator. For each maximum of
the gauge functional $F_U[g]$, besides of its 
discarded eight trivial zero modes, the eigenvalues of the \mbox{F-P} operator 
are positive. If all $\lambda_i>0$ the configuration is said to be
within the Gribov horizon.   

Having the lowest $n$ eigenvalues and eigenvectors determined,
one can construct an approximation to the ghost propagator 
\begin{equation}
  G_n(k) = \sum_{i=1}^{n}
  \frac{1}{\lambda_i}\,\vec{\Phi}_i(k)\cdot\vec{\Phi}_i(-k) 
\label{eq:Def-ghost-q-by-spectrum}
\end{equation}
with $\vec{\Phi}_i(k)$ being the Fourier transform of the $i$-th eigenmode. 
If all $n=8V$ eigenvalues and eigenvectors were known, 
the ghost propagator $G$ would be completely determined.
In the recent literature~\cite{Greensite:2004ur,Gattnar:2004bf} it is
stated that the \mbox{F-P} operator acquires very small eigenvalues in the
presence of vortex excitations. In any case, according to a
popular belief it is an enhanced density of eigenvalues near zero 
which causes the ghost propagator to diverge stronger
than $1/q^2$ near zero momentum \cite{Zwanziger:2003cf}.

At $\beta=6.2$ we have extracted the 200 (50) lowest (non-trival)
eigenvalues and their corresponding eigenfunctions on a $12^4$ and
$16^4$ ($24^4$) lattice. Additionally, 90 eigenvalues have been
extracted on a $24^4$ lattice at $\beta=5.8$. This allows us to check
that low-lying eigenvalues are shifted towards $\lambda=0$ as the
physical volume is increased.  In order to clarify how the choice of
gauge copy influences the spectrum, the eigenvalues and eigenvectors
have been extracted separately on our sets of \fc{} and \bc{}
gauge-fixed configurations.

In \Fig{fig:fps_lowest} the distributions of the
lowest $\lambda_1$ and second lowest $\lambda_2$ eigenvalues of the 
\mbox{F-P} operator are shown for different (physical) volumes.
There $h(\lambda,\lambda+\Delta)$ represents the
average number (per configuration) of eigenvalues 
found in the intervall $[\lambda,\lambda+\Delta]$. 
Open bars refer to the distribution on \fc{} gauge copies
while full bars to that on \bc{} copies.
From this figure it is quite obvious that both
eigenvalues, $\lambda_1$ and $\lambda_2$, are shifted to lower values as
the physical volume is increased. In conjunction the spread of
$\lambda$ values is decreased. We have found that the center of those
distributions tends towards zero stronger than $ 1/L^{2}$. Here $L$
refers to the linear lattice extension in physical units. For example,
we have found 
$\langle\lambda_{1}\rangle(L)\propto 1/L^{2+\varepsilon}$ with
$\varepsilon\approx0.16(4)$. It is also visible that 
the eigenvalues $\lambda_1$ and $\lambda_2$ on \fc{} gauge
copies are on average lower than those obtained on
\bc{} copies. 

\Eq{eq:Def-ghost-q-by-spectrum} suggests the popular belief 
that low-lying eigenvalues and eigenvectors have a major impact on the ghost 
propagator at the lowest momenta. Actually, this is not so easy to predict.
We have studied, to what extent the lowest $n$ modes saturate the estimator 
of the ghost propagator $G(k)$ on a given set of gauge copies. 
We show the result in \Fig{fig:ghfps_n} for the lowest ($q_1$) and the second 
lowest momentum ($q_2$) available on lattices of different size at $\beta=6.2$. 
The estimates for $G_n(k)$ have been divided by the
full (conjugate gradient) estimator $G(k)$ in order to compare the 
approach to saturation for different volumes. 

\begin{floatingfigure}[l]
  \includegraphics[width=7cm]{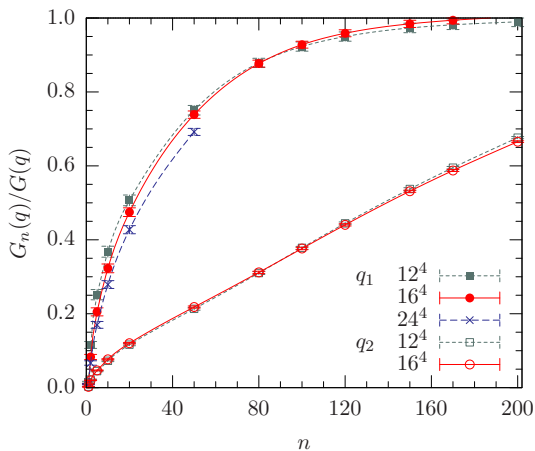}
  \caption{The ratio of the ghost
    propagator $G_n(q)$, approximated by the $n$ 
    lowest \mbox{F-P} eigenmodes and eigenvalues, to the corresponding
    full estimator $G(q)$ is shown as a function of $n$ 
    for the lowest ($q_1$) 
    and second lowest momentum ($q_2$). The inverse coupling 
    is $\beta=6.2$
    and the lattice size ranges from $12^4$ to $24^4$.}
  \label{fig:ghfps_n}
\end{floatingfigure}

Considering first the lowest momentum $q_1$ we observe from
\Fig{fig:ghfps_n} that the rates of convergence differ,
albeit slightly, for the three different lattice sizes. The relative
deficit rises with the lattice volume. For the $12^4$ and $16^4$
lattice the rates of convergence are still about the same. 
For example, taking only 20 eigenmodes is definitely too less. One is short of  
reproducing the ghost propagator (by about 50\%) whereas
$150\ldots200$ eigenmodes are sufficient to 
reproduce it within a few percent. 
For the second lowest momentum $q_2$ even 200 eigenmodes are far from
saturating the result.

In addition, the inverse participation ratio has been calculated 
for the modes in order to figure out the localization properties of 
the \mbox{F-P} eigenmodes. It turns out that the lowest eigenmodes are not
localized on average, although there are, albeit few, relatively
localized ones.

\section{Conclusions}

We have studied the lower momentum region of Landau gauge
gluodynamics using Monte Carlo simulations on lattice sizes
$16^4$, $24^4$ and $32^4$. The inverse bare coupling constant has been
set to $\beta=5.8$, $6.0$ and $6.2$ using the Wilson action.

Concerning the influence of Gribov copies, it turns out that 
the gluon propagator does not depend on the chosen gauge 
copy while the ghost propagator does. There are indications, however, 
that the strength of this influence, at the same physical momentum,
decreases with increasing volume. Towards $q \to 0$ in the infrared
momentum region, the gluon dressing function is decreasing, while the ghost 
dressing function is increasing. However, the power laws predicted
from the DSE approach cannot be confirmed from our data.
Correspondingly, the behavior of the running coupling $\alpha_s(q^2)$ 
in a suitable momentum subtraction scheme (based on the
ghost-ghost-gluon vertex) does not approach the expected finite limit. 
Instead, this coupling has been found to decrease for lower momenta 
after passing a turnover at $q^2\approx0.4$~GeV$^2$. 

In addition, we have investigated the spectral properties of the \mbox{F-P}
operator and its relation to the ghost propagator 
on a $24^4$ lattice at $\beta=5.8$ and on $12^4$, $16^4$ and $24^4$ lattices 
at $\beta=6.2$. As expected from another study \cite{Zwanziger:2003cf} we
have found that the low-lying eigenvalues are shifted towards zero
(the configurations approaching the Gribov horizon) as the volume is increased. 
The low-lying eigenvalues extracted on \bc{} gauge copies (those with the
largest functional value) are larger on average than those on 
\fc{} copies. Thus, for finite volumes better gauge-fixing (in terms of
the gauge functional) keeps the configurations at larger distance from
the Gribov horizon. The study of the contributions to the ghost
propagator coming from the lowest  
eigenmodes of the \mbox{F-P} operator reveals that even the ghost propagator
at the lowest momentum needs not less than $150 \ldots 200$ modes for 
being sufficiently approximated. This result refers to our smallest ($12^4$)
lattice at $\beta=6.2$. For larger volumes and higher lattice momenta
the amount of modes needed in an eigenmode expansion is even larger.

\bigskip
{\small
All simulations have been done on the IBM pSeries 690 at HLRN.  
We thank Christian S. Fischer, Lorenz von Smekal and Anthony G. Williams
for inspiring discussions. We are grateful to Hinnerk St\"uben for
contributing parts of the program code.}


\bibliographystyle{JHEP}

\providecommand{\href}[2]{#2}\begingroup\raggedright\endgroup

\end{document}